\documentclass[a4paper]{jpconf}
\usepackage{graphicx}
\usepackage[square,sort&compress,numbers]{natbib}

\begin{document}

\title{A pressurized argon gas TPC as DUNE near detector}
\author{J Mart\'in-Albo, for the DUNE Collaboration}
\address{Department of Physics, University of Oxford, United Kingdom}
\ead{justo.martin-albo@physics.ox.ac.uk}

\begin{abstract}
DUNE is a new international experiment for neutrino physics and nucleon decay searches. It will consist of two detectors, about 1300 km apart, exposed to a multi-megawatt neutrino beam that will be built at Fermilab. One of the two detectors will be installed several hundred meters downstream of the neutrino production point with the primary role of characterising the energy spectrum and composition of the beam as well as performing precision measurements of neutrino cross sections. For the design of this so-called near detector, the DUNE Collaboration is considering, among other technologies, a pressurized argon gas time projection chamber. Such a detector, thanks to its low density and low detection thresholds, would allow the detailed measurement in argon of nuclear effects at the neutrino interaction vertex, which are considered at present one of the most important sources of systematic uncertainty for neutrino oscillation measurements.
\end{abstract}

\section{The DUNE project}
The \emph{Deep Underground Neutrino Experiment} (DUNE) \cite{Acciarri:2016crz, Acciarri:2015uup, Acciarri:2016ooe} is an international effort to build a next-generation long-baseline oscillation experiment between Fermilab (Illinois, USA), where a new megawatt-scale neutrino beamline \cite{Strait:2016mof} will be built, and a 40-kilotonne liquid argon detector installed at the Sanford Underground Research Facility (South Dakota, USA), about 1300~km away. The main scientific objective of DUNE is the precision measurement of neutrino oscillations, including testing CP violation in the lepton sector and determining the neutrino mass ordering. Besides, the massive DUNE far detector, consisting of four 10-kt liquid argon (LAr) TPC modules located 1~km underground, will offer unique capabilities for addressing additional non-accelerator physics topics such as proton-decay searches or the detection of the neutrino burst from core-collapse supernovae.

To meet the ultimate systematic precision needed to fulfil the DUNE physics goals, another detector will be installed about 600 meters away from the start of the neutrino beamline. This so-called \emph{near detector} must precisely characterize the neutrino beam energy and composition and measure to unprecedented accuracy the cross sections and particle yields of the various neutrino scattering processes. As the near detector will be exposed to an intense flux of neutrinos, it will collect an extraordinarily large sample of neutrino interactions, allowing for an extended science program, including searches for heavy sterile neutrinos or non-standard interactions. 

\section{Motivations for an argon gas TPC as near detector}
\begin{figure}
\centering
\includegraphics[width=0.525\textwidth]{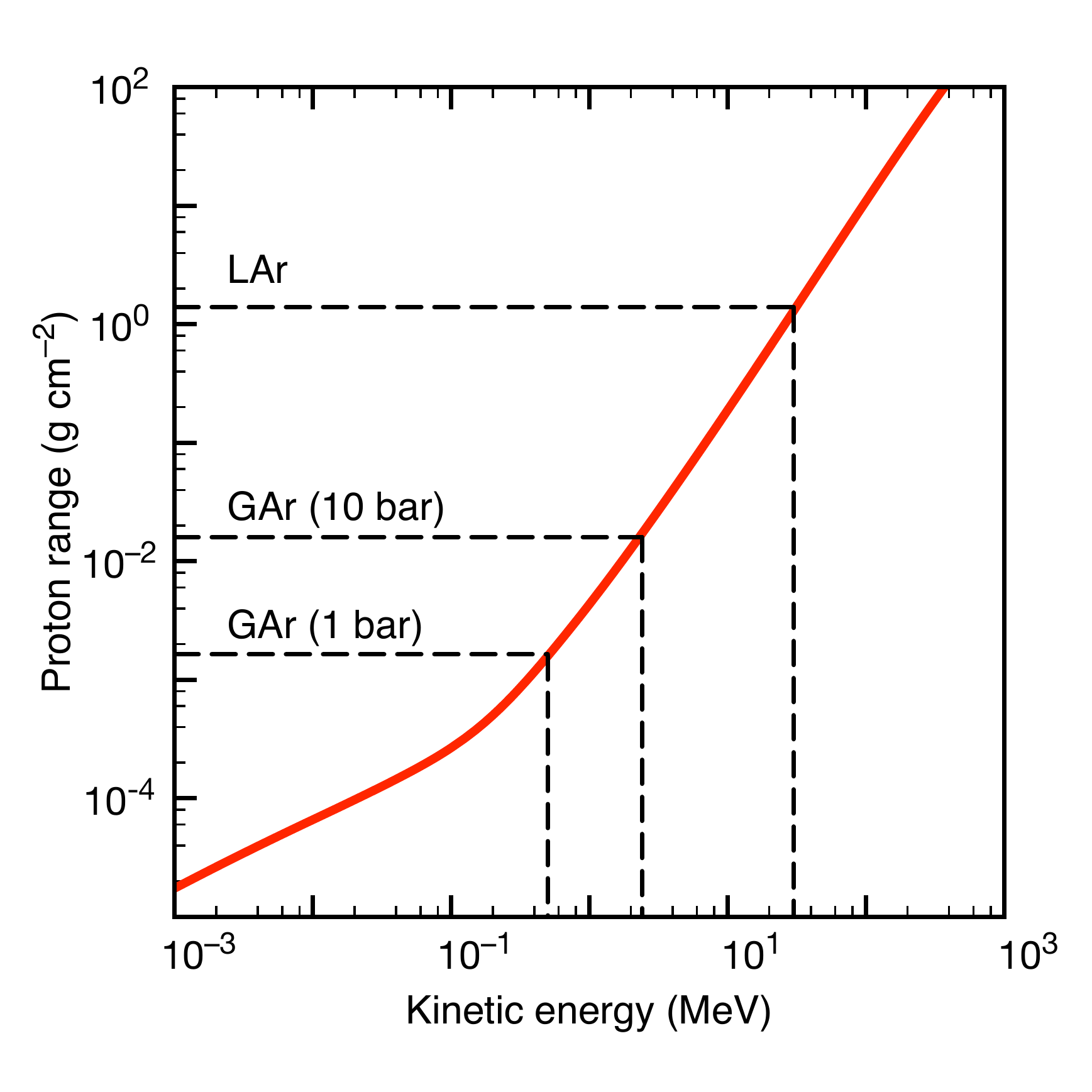}
\caption{Predicted range (CDSA) for protons in argon as a function of their kinetic energy. The dashed lines show the resulting energy detection thresholds for 1-centimetre-long proton tracks and three different argon densities.}
\label{fig:ProtonRangeArgon} 
\end{figure}

Oscillation measurements require a thorough understanding of neutrino interactions to accurately reconstruct the energy of the incoming neutrino and predict the flux at the far detector from the measurements at the near detector. Experiments rely on nuclear models to relate the near-detector measurements to the initial neutrino energy and spectra. However, much of our understanding of neutrino scattering comes from light-nuclei data and hence the models probably do not accurately represent the physics of heavier targets such as argon (the target nucleus of the DUNE far detector), where nuclear effects like nucleon correlations and final state interactions introduce significant complications and biases that result in large systematics. Additional experimental data are therefore required to clarify these issues and improve the models describing neutrino-nucleus interactions. An argon gas TPC offers unique capabilities for such a task: it can provide a high-resolution measurement of the charged tracks emitted from the neutrino interaction vertex, including, thanks to its low detection thresholds (see Figure~\ref{fig:ProtonRangeArgon} for a comparison with LAr), those in the low energy region in which the differences between model predictions are most pronounced \cite{Hamilton:2015ows}.

\section{Detector concept}
\begin{figure}
\centering
\includegraphics[width=0.8\textwidth]{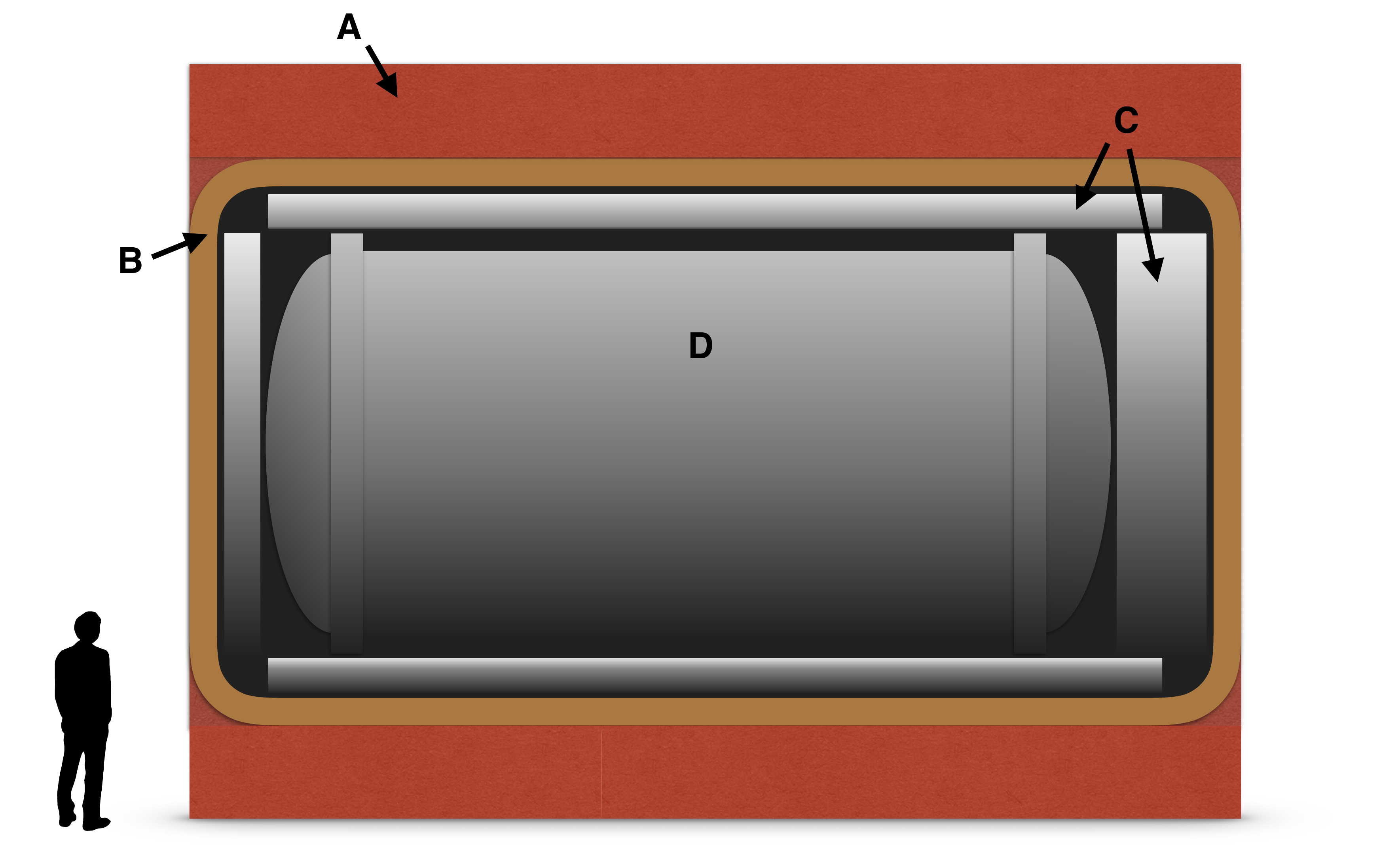}
\caption{Schematic drawing of a possible design of the DUNE near detector based on an argon gas TPC. The main components of the detector are labelled: A) magnet return yoke; B) magnet coil; C) electromagnetic calorimeters; and D) pressure vessel and TPC.} \label{fig:GArTPC-ND}
\end{figure}

Figure~\ref{fig:GArTPC-ND} shows a possible near detector design based on an argon gas TPC. It consists of the following elements:
\begin{itemize}
\item A large \emph{time projection chamber} (TPC) of square cross section of about $2.5\times2.5$~m$^2$ with an electric drift field perpendicular to the neutrino beam direction and parallel to the floor. The TPC offers low detection thresholds, excellent tracking performance (point resolution below 1 mm and two-track separation better than 15 mm), high-resolution momentum measurement ($<5\%$ for 1 GeV tracks) and particle identification based on d$E$/d$x$.
\item A cylindrical \emph{pressure vessel} of $\sim65$~m$^3$ that houses the TPC and can hold about 1 tonne or argon at 10~bar. To minimize the inactive mass in the near detector, the vessel could be manufactured with either light alloys (e.g. titanium or aluminium) or composite materials.
\item An \emph{electromagnetic calorimeter} made of layers of lead and plastic scintillator that surrounds the TPC and detects the neutral particles that leave its active volume.
\item A \emph{dipole magnet} that surrounds the entire detector establishing a uniform magnetic field of 0.4~T perpendicular to the neutrino beam. The separation between fluxes of neutrinos and antineutrinos requires a magnetized neutrino detector to charge-discriminate the leptons and antileptons produced in the neutrino charged-current interactions.
\end{itemize}

\section{Outlook}
The DUNE Collaboration is considering several technologies for the design of the experiment's near detector. Among them, an argon gas TPC, which offers unique capabilities for the measurement of nuclear effects in neutrino interactions, one of the most significant sources of systematic uncertainty for neutrino oscillation experiments. R\&D programs in both Europe and the US will address the most significant design challenges of the detector (e.g.\ read-out technology or gas quenchers).

\ack
This work has been supported by the Science and Technology Facilities Council (STFC) of the United Kingdom.


\bibliographystyle{iopart-num}
\bibliography{references.bib}

\providecommand{\newblock}{}
\begin{thebibliography}{1}
\expandafter\ifx\csname url\endcsname\relax
  \def\url#1{{\tt #1}}\fi
\expandafter\ifx\csname urlprefix\endcsname\relax\def\urlprefix{URL }\fi
\providecommand{\eprint}[2][]{\url{#2}}

\bibitem{Acciarri:2016crz}
Acciarri R {\em et~al.\/} (DUNE Collaboration) 2016  (\textit{Preprint}
  \eprint{1601.05471})

\bibitem{Acciarri:2015uup}
Acciarri R {\em et~al.\/} (DUNE Collaboration) 2015  (\textit{Preprint}
  \eprint{1512.06148})

\bibitem{Acciarri:2016ooe}
Acciarri R {\em et~al.\/} (DUNE Collaboration) 2016  (\textit{Preprint}
  \eprint{1601.02984})

\bibitem{Strait:2016mof}
Strait J {\em et~al.\/} 2016  (\textit{Preprint} \eprint{1601.05823})

\bibitem{Hamilton:2015ows}
Hamilton P~A 2015 {\em {A study of neutrino interactions in argon gas}\/} Ph.D.
  thesis Imperial College London
  \urlprefix\url{http://www.t2k.org/docs/thesis/062}

\end{thebibliography}

\end{document}